\documentclass[journal=jacsat,manuscript=article]{achemso}

\usepackage[version=4]{mhchem}        
\usepackage{fix-cm}
\usepackage{graphicx}                 
\usepackage{amsmath,amssymb}          
\usepackage{physics}                  
\usepackage{xspace}                   
\usepackage{makecell}                 
\usepackage{tabularx}                 
\usepackage{array}                    
\usepackage{epstopdf}                 
\usepackage{xfrac}
\usepackage{soul}    
\usepackage{xcolor}
\usepackage{float}

\usepackage[
    colorlinks = true,
    linkcolor = blue,
    urlcolor  = blue,
    citecolor = blue,
    anchorcolor = blue
]{hyperref}

\newcommand{\Efield}{$E$-field\xspace}

\begingroup
\small

\author{Niccol\'{o} Fontana}
\affiliation{Department of Physics, University of Oxford, The Clarendon Laboratory, Parks Road, Oxford OX1 3PU, UK}

\author{Mikhail V. Vaganov}
\affiliation{Department of Physics, University of Oxford, The Clarendon Laboratory, Parks Road, Oxford OX1 3PU, UK}

\author{Gabriel Moise}
\affiliation{Department of Physics, University of Oxford, The Clarendon Laboratory, Parks Road, Oxford OX1 3PU, UK}

\author{William K. Myers}
\affiliation{CAESR, Inorganic Chemistry Laboratory, University of Oxford, South Parks Road, Oxford OX1 3QR, UK}

\author{Kun Peng}
\affiliation{Department of Physics, University of Oxford, The Clarendon Laboratory, Parks Road, Oxford OX1 3PU, UK}

\author{Arzhang Ardavan}
\affiliation{Department of Physics, University of Oxford, The Clarendon Laboratory, Parks Road, Oxford OX1 3PU, UK}
\email{arzhang.ardavan@physics.ox.ac.uk}

\author{Junjie Liu}
\affiliation{Department of Physics, University of Oxford, The Clarendon Laboratory, Parks Road, Oxford OX1 3PU, UK}
\alsoaffiliation{School of Physical and Chemical Sciences, Queen Mary University of London, London E1 4NS, UK}
\email{junjie.liu@qmul.ac.uk}

\endgroup

\title{\Large Electric-field Quantum Sensing Exploiting a Photogenerated Charge-transfer Triplet State in an Organic Molecule}

\begin{document}

\begin{abstract}
Molecular spin systems are promising platforms for quantum sensing due to their chemically tunable Hamiltonians, enabling tailored coherence properties and interactions with external fields. However, electric field sensing remains challenging owing to typically weak spin-electric coupling (SEC) and limited directional sensitivity. Addressing these issues using heavy atoms exhibiting strong atomic spin-orbit couplings (SOC) often compromises spin coherence times. Here, we demonstrate coherent electric field sensing using a photogenerated charge-transfer (CT) spin triplet state in the organic molecule ACRSA (10-phenyl-10H,10\textquotesingle H-spiro[acridine-9,9\textquotesingle-anthracen]-10\textquotesingle-one). By embedding electric field pulses within a Hahn echo sequence, we coherently manipulate the spin triplet and extract both the magnitude and directional dependence of its SEC. The measured SEC strength is approximately 0.51 Hz/(V/m), comparable to values reported in systems with strong atomic SOC, illustrating that heavy atoms are not a prerequisite for electric-field sensitivity of spin states. Our findings position organic CT triplets as chemically versatile and directionally sensitive quantum sensors of \(E\)-fields that function without atomic-SOC-mediated mechanisms.
\end{abstract}

\begin{section}{Introduction}

Quantum sensing exploits the unique properties of quantum systems such as quantum coherence and entanglement to achieve unprecedented precision in measuring physical quantities~\cite{qsensing}. This paradigm has enabled remarkable advances across diverse platforms, including nitrogen-vacancy centres in diamond~\cite{NV}, superconducting circuits~\cite{SC}, and cold atoms~\cite{CA}. Quantum sensing of electric fields, in particular, has seen rapid development through systems like Rydberg atoms~\cite{rydberg}, trapped ions~\cite{ion_traps}, and superconducting circuits~\cite{supercond}, which offer excellent sensitivity \textit{via} strong coupling to electric fields. Yet, these platforms typically operate at micrometer to millimeter scales, limiting their spatial resolution and hindering their ability to access electric fields near surfaces or within heterogeneous environments~\cite{sens1, sens2}.

\medskip

\noindent Molecular electron spins have long been studied as promising candidates for magnetic field sensing due to their well-defined spin states and intrinsic coupling to magnetic fields \textit{via} the Zeeman effect~\cite{MCC1}. Another strength lies in the ability to extensively tailor their properties through chemical design, enabling controlled coupling to electric~\cite{HoW10}, optical~\cite{Opt}, and mechanical~\cite{Mech} degrees of freedom, as well as impressively long phase coherence times, up to milliseconds at 10 K~\cite{long_T2}. Building on these advantages, recent efforts have begun exploring molecular spins for electric field sensing. Molecular systems offer a compelling route, as their intrinsically nanoscale dimensions enable high spatial resolution and placement in direct proximity to the sensing target~\cite{DoF}. 

\medskip

\begin{figure}
        \includegraphics[width = 0.75\linewidth, height = 8cm]{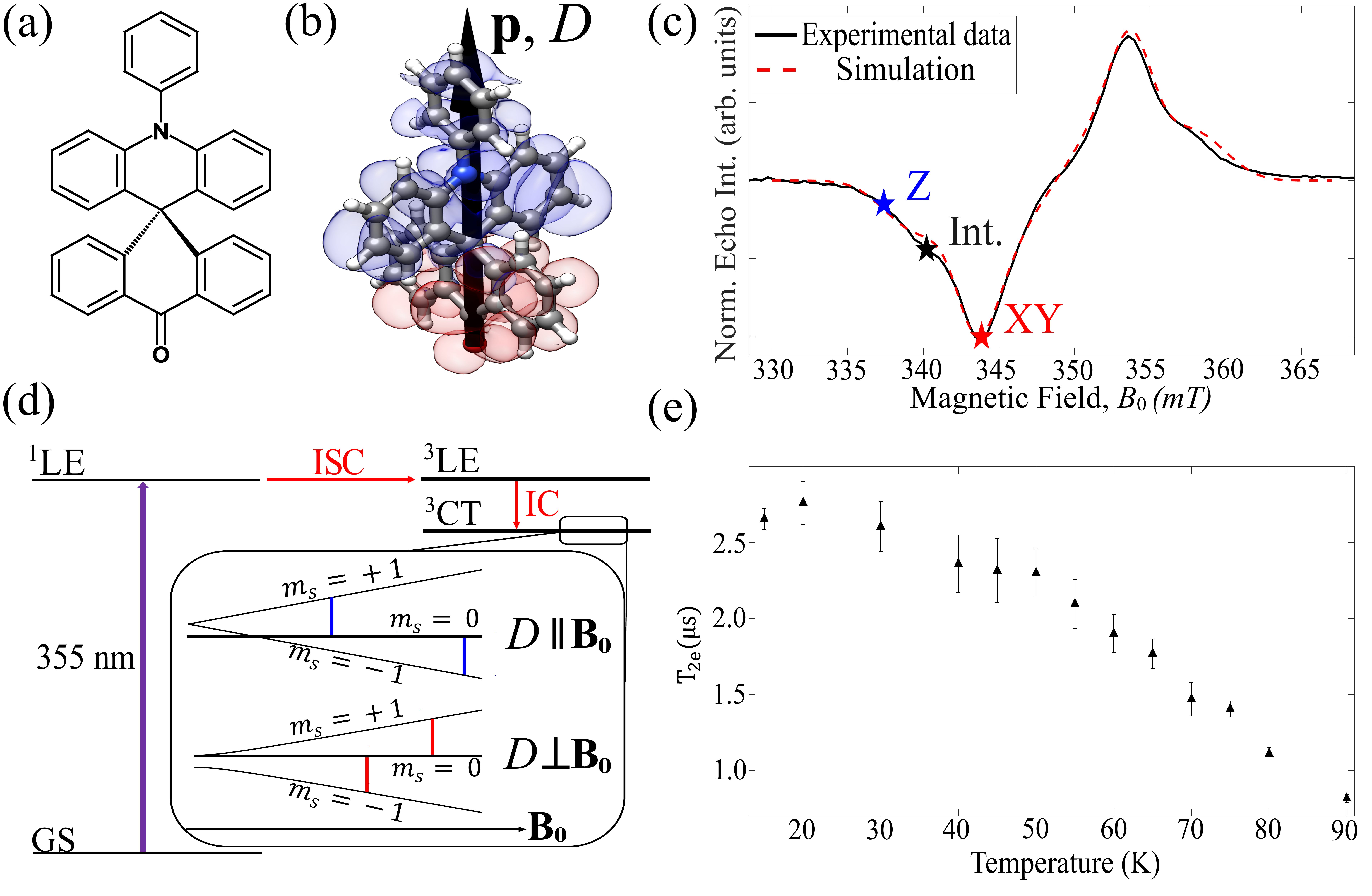}
        \caption{
    (a) The schematic molecular structure of ACRSA.
    (b) The spin density of the charge-transfer triplet state induced by photoexcitation with $\mathrm{\lambda}$ = 355 nm, computed with the ORCA software (B3LYP/EPR-II basis). The blue (red) molecular orbitals correspond to the lower (higher) singly occupied molecular orbitals, coinciding with the hole (electron) density. The black arrow is the predicted orientation of the longitudinal zero-field splitting tensor/electric dipole moment, and it is defined as the $z$-axis for the experiment.
    (c) The (X-band) experimental and simulated field sweep of ACRSA doped into a PMMA matrix ($\sim 90$ $\mu$M) at 20 K. The three starred points correspond to the fields where we conducted the spin-electric coupling measurements. 
    (d) Schematic showing the optical pathway that leads to the formation of the triplet charge-transfer state to be electrically modulated, including the initial photoexcitation with $\lambda = 355$~nm, intersystem crossings (ISC), and internal conversion (IC). Inset: The simplified Zeeman energy diagrams when the longitudinal zero-field splitting ($D$) is parallel (top) and perpendicular (bottom) to the principal magnetic field, together with the allowed EPR transitions. The light-induced triplet state is spin polarised with $\sim 95\%$ population of $|m_s = 0\rangle$ (indicated by the thick lines).
    (e) The electronic phase-memory time ($T_{2\mathrm{e}}$) as a function of temperature. $T_{2\mathrm{e}}$ exceeds 2.5 $\mu$s at 20~K, and remains above 1.0 $\mu$s at 77~K. 
    }
    \label{fig: Intro}
\end{figure}

\noindent Progress toward electric field sensing with molecular systems has revealed that molecular spins can exhibit strong coupling to electric fields~\cite{EFC, Kintzel2021, SEC2, Vaganov2025, Cini2025, Liu2025}. However, to achieve sufficient sensitivity using electron spin resonance, a significant spin polarisation in magnetically diluted samples is desirable, often necessitating low temperatures. An alternative route to large spin polarisations is offered by light-induced species such as spin-correlated radical pairs (SCRPs), which have shown room-temperature operation, owing to the generation of polarised spin states through spin-selective processes during their formation~\cite{SCRP1}. SCRPs have been widely investigated as potential candidates for spin qubits, particularly for implementing 2-qubit gates, where two electron spins are correlated \textit{via} zero-field splitting or J-coupling~\cite{DoF}. Quantum teleportation has already been successfully demonstrated using SCRPs, highlighting their ability to embody entangled states between two electron spins within a molecule \cite{SCRP2}. Recently, Xie \textit{et al.} \cite{SCRP3} demonstrated the use of SCRPs for electric-field sensing by encapsulating a cyclophane host around one of the radical pair partners. In this case, the local supramolecular electric field modulates the inter-spin distance and leads to a different modulation frequency in the out-of-phase component of the spin echo. Crucially, however, such electric field effects on light-induced spin states have, until now, not been demonstrated using externally applied (and controllable) \(E\)-fields.

\medskip

\noindent Here, we report the spin-electric coupling (SEC) in a commercially available organic molecular semiconductor spiro-acridine-anthracenone known as ACRSA (10-phenyl-10H,10\textquotesingle H-spiro\newline
[acridine-9,9\textquotesingle-anthracen]-10\textquotesingle-one). Our investigation does not involve structural modifications to the molecule, but instead focuses on detecting externally generated electric fields (and their direction), which alter the system's resonance frequency. This frequency shift can be measured using the modified Electron Paramagnetic Resonance (EPR) spin-echo sequence first proposed by Mims \cite{Mims}. Facilitating practical implementation, ACRSA molecules can be doped into poly(methyl methacrylate) (PMMA) polymer thin films and incorporated into device architectures suitable for measuring transient and/or alternating electric fields (see Fig. S4 in the Supporting Information). 

\medskip

\begin{figure}[H]
    \centering
        \includegraphics[width=\linewidth]{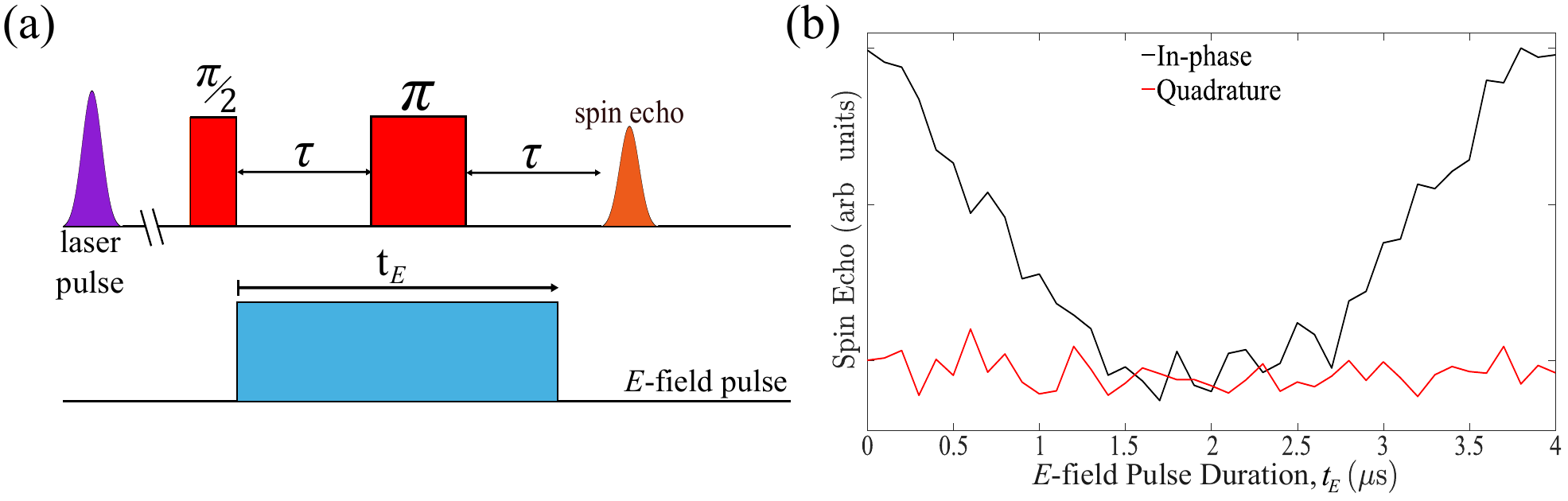}
        \caption{%
        (a) The modified Hahn echo sequence for SEC measurements.  A laser pulse at 355 nm generates the $\mathrm{^3CT}$ state with a spin-polarised initial population. After a fixed delay, a Hahn-echo sequence measures the spin coherence of the $\mathrm{^3CT}$ state. An $E$-field pulse is inserted immediately after the $\pi/2$ microwave pulse and the echo signal is recorded as a function of the duration and/or amplitude of the $E$-field pulse.
        (b) The echo intensity as a function of the $E$-field pulse duration, t$_E$. The data were recorded at 20 K with $\tau = 2$ $\mu$s and $B_0$ at the ``Int.''\ field-position (as indicated in Fig.~\ref{fig: Intro}(c)) with an \(E\)-field of 1.5 \(\times 10^6\) V/m. The absence of an electric field response in the quadrature channel arises from the combination of a linear spin–electric coupling and the random orientation of spins within the ensemble.}
    \label{fig: PulseLin}
\end{figure}

\noindent We attribute the observed SEC in ACRSA to the significant electric dipole associated with its charge-transfer state. Although organic molecules typically lack strong atomic spin-orbit coupling (SOC), a feature often deemed essential for enhancing the SEC \cite{HoW10}, the SEC observed here is comparable in magnitude to that reported in transition-metal-based systems~\cite{EFC, SEC2}. Notably, weak SOC is also associated with longer spin coherence times~\cite{EFM}, even at elevated temperatures. This property enhances the sensitivity of electric field quantum sensing using SCRPs at room temperature.

\end{section}

\begin{section}{Results and Discussion}
\begin{subsection}{Optical and Magnetic Properties of ACRSA}

ACRSA has been extensively studied in the context of organic LEDs \cite{acrsa_0}, specifically as an efficient thermally activated delayed ﬂuorescence material, due to its strong reverse intersystem crossing (rISC) \cite{acrsa_2}. The molecular structure, shown in Fig. \ref{fig: Intro}(a), consists of an electron-donating acridine unit and an electron-accepting anthracenone moiety, connected \textit{via} a spiro-junction. This orthogonal arrangement results in weak coupling between the two $\pi$-systems, due to their spatial separation and limited orbital overlap. Both theoretical \cite{acrsa_1} and experimental \cite{Franca2023} investigations have characterised the electronic structure and photophysics of ACRSA as a function of the solvent and the excitation wavelength. As highlighted in \cite{acrsa_2}, upon excitation at $\lambda = 355$ $\mathrm{nm}$, a vibronically-assisted optical transition occurs between the singlet ($^1A_1$, $C_{2v}$ group) ground state and an excited singlet state localised on the anthracenone moiety ($\mathrm{^1LE (2 ^1A_1)_{n \pi^*}}$). This is followed by a cascade of radiationless ISC processes to the localized triplet state ($\mathrm{^3LE(1 ^3A_2)_{A \pi \pi^*}}$), which then undergoes internal conversion to a charge-transfer triplet state ($\mathrm{^3CT(2 ^3A_2)_{\pi \pi^*}}$). A summary of these processes is shown in the Jablonski diagram in Fig. \ref{fig: Intro}(d). 

\medskip

\noindent Fig. \ref{fig: Intro}(b) shows the spin distribution of the photogenerated electron-hole pair, as calculated with DFT in the B3LYP/EPR-II basis using the ORCA software \cite{Orca}. The blue surface denotes the lower energy singly occupied molecular orbital (SOMO), representing the hole density, while the red surface represents the higher SOMO. Importantly, this electron-hole pair gives rise to both a charge and a spin separation. The former results in an electric dipole moment $\mathbf{p}$ ($\sim 23$ Debye from the DFT calculations), while the latter leads to a zero-field splitting in the spin states characterised by an (almost perfectly) axially symmetric $\mathbf{D}$ tensor. Both $\mathbf{p}$ and the magnetic anisotropy axis (assuming a uniaxial symmetry) are predicted to be closely aligned with the molecular $z$-axis, defined as the direction connecting the nitrogen and the oxygen atoms in the molecular structure, as shown in Fig. \ref{fig: Intro}(b). It is important to highlight that $\mathbf{p}$ and $\mathbf{D}$ both depend on the electron-hole pair wavefunction, providing an essential link between the magnetic and the electric degrees of freedom in this molecule. 

\medskip

\begin{figure}[H]
    \centering
    \includegraphics[width = \linewidth]{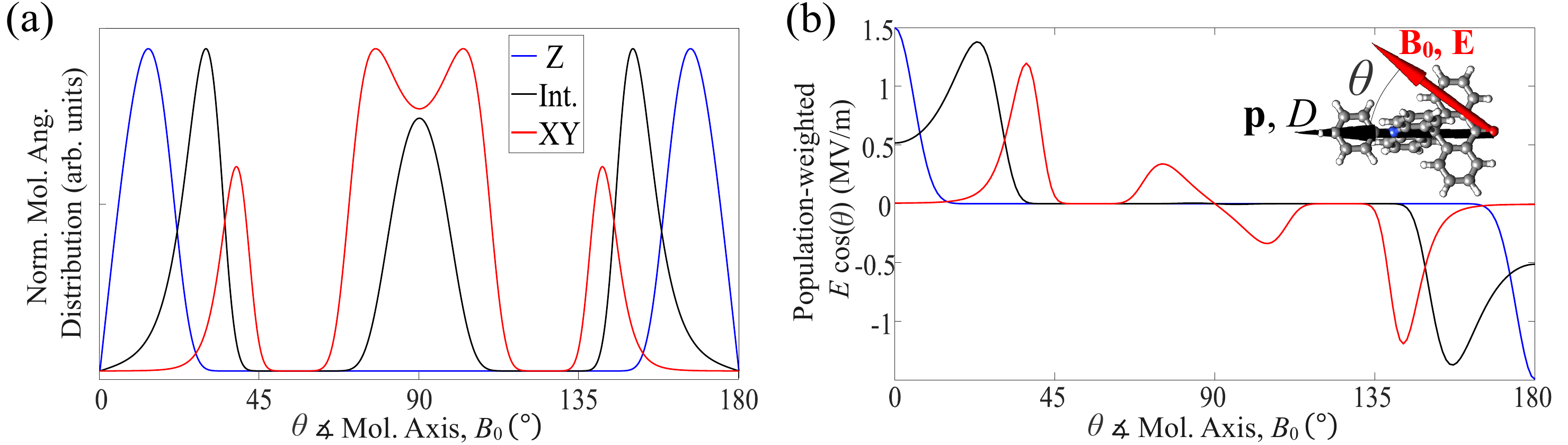}
    \caption{
    (a) (Simulated) Normalized angular distribution of resonant molecules as a function of the angle \(\theta\) between the external magnetic field \(\mathbf{B_0}\) and the molecular dipole/orientation axis \textbf{p}/\(D\), as illustrated in the inset of panel (b). The three distributions correspond to the magnetic-field values used in the SEC study (see Fig. \ref{fig: Intro}(c)). The observed peaks in the XY and Int. distributions arise from the angular dependence of the resonant field, as detailed in Fig. S3 in the Supporting Information. (b) (Simulated) Effective electric field, calculated as the projection of the electric field along the \textbf{p}/\(D\) axis (\(E \cdot \cos(\theta)\), with \(E = \) 1.5 \(\mathrm{\sfrac{MV}{m}}\)) weighted by the \(\mathbf{B_0}\)-dependent molecular population shown in (a). This represents the strength of the interaction between the electric field and the molecular ensemble, which takes into account the angular distribution variations at different \(B_0\).}
    \label{fig: Distr}
\end{figure}

\noindent Fig. \ref{fig: Intro}(c, solid line) shows the X-band echo-detected field-swept (EDFS) EPR spectrum measured on an ensemble of randomly oriented ACRSA diluted in a matrix of PMMA at 20~K, following $\mathrm{\lambda = 355}$ nm excitation. The EDFS spectrum of the $\mathrm{^3CT}$ state can be simulated using the EasySpin software package \cite{EasySpin} with the following spin Hamiltonian
\begin{equation}
    \mathcal{\hat{H}} = 
    D\hat{S}_z^2 + \mu_B g_e \mathbf{B_0} \cdot \hat{\mathbf{S}},
     \label{Spin_Ham}
\end{equation}
with $S = 1$, an isotropic $g_e = 2.0$ and a uniaxial anisotropy of $|D| = 317 \pm 12$~MHz. Assuming an easy-plane type anistropy with $D > 0$ (in agreement with DFT calculations), the simulation suggests the light-induced initial state at zero field is with 95\% $|m_s = 0\rangle$, 2.5\% $|m_s = +1\rangle$ and 2.5\% $|m_s = -1\rangle$. The population difference between the $|m_s = 0\rangle$ and $|m_s = \pm1\rangle$ spin sublevels in ACRSA is approximately $90\%$, indicating a nearly complete spin polarisation. This value is significantly higher than the thermal population differences typically observed between the ground and excited states in molecular magnets exhibiting a sizable SEC. For example, the population difference is only $1.7\%$ for the $S = 1$ antiferromagnetic ring \cite{EFC} and ranges from $1\%$ to $2.8\%$ for the $S = 5/2$ Mn-based molecules\cite{Vaganov2025} at 20~K. The combination of EPR data and DFT simulations suggests that the magnetic anisotropy of the $\mathrm{^3CT}$ state arises from the magnetic dipole interaction between the spatially separated electron–hole pair, with negligible contribution from atomic SOC. This is expected in organic molecules without heavy atoms in their structure, as in the case of ACRSA \cite{heavy_atoms}.

\medskip

\noindent To assess the coherence properties of the \(\mathrm{^3CT}\) state, we measured the temperature dependence of the ACRSA electron spin phase-memory time \(T_{2\mathrm{e}}\), as shown in Fig. \ref{fig: Intro}(e). At liquid nitrogen temperature, the molecule exhibits a \(T_{2\mathrm{e}}\) of approximately \( 1~\mu\text{s}\). As the temperature decreases, \(T_{2\mathrm{e}}\) increases steadily, eventually saturating at about \(2.5~\mu\text{s}\) at \(T = 20~\text{K}\). In comparison, the lifetime of the light-induced \(\mathrm{^3CT}\) state is reported to approach 200~$\mu$s at 20~K\cite{acrsa_2} (see Fig. S1 Supporting Information), significantly longer than the observed spin coherence time. This indicates that the triplet-state lifetime does not constrain the spin dynamics.

\medskip

\noindent Because the magnetic anisotropy ($D$) of the light-induced state is directly related to the dipole interaction between the electron and hole, we anticipate an external electric field to modulate $D$ \textit{via} coupling to the molecular electric dipole ($\mathbf{p}$). The electric field influences the interaction between the electron-hold pair by modulating the electronic structure and/or the geometry of the molecule, which manifests as an electric-field dependence of $D$. Owing to the lack of inversion symmetry for the electronic structure of the light-induced state, a linear SEC effect is expected to first order, that is, the $E$-field-induced modulation of $D$, $\delta D(E)$, is given by
\begin{equation}
     \delta D(E)  = \kappa E\cdot\cos(\theta),
     \label{eq: dDz}
\end{equation}
where $E$ is the external $E$-field, $\kappa$ is the SEC coupling coefficient and $\theta$ represents the angle between the external $E$-field and $\mathbf{p}$. To simplify the model, we make the approximation that both $\mathbf{D}$ and $\mathbf{p}$ are collinear with the molecular $z$-axis. We test this hypothesis by measuring the $E$-field-induced modulation of the spin echo signal in pulsed EPR experiments.

\end{subsection}

\begin{subsection}{Spin-Electric Coupling Measurement and Theoretical Model}
We investigated the SEC in ACRSA using the modified Hahn echo sequence in Fig. \ref{fig: PulseLin}(a), comprising an initial laser pulse at a wavelength of 355 nm which generates the $\mathrm{^3CT}$, followed by a Hahn-echo sequence measuring the spin echo signal. A square DC $E$-field pulse is applied immediately after the $\pi/2$ microwave pulse and the echo signal is recorded as a function of the duration/amplitude of the $E$-field pulse. The presence of a SEC leads to a change in the spin energy of ACRSA, and consequently alters the spin transition frequency. The in-phase component of the spin echo signal is expected to follow a $\cos{(2\pi\delta f\times t_E)}$ behaviour, where $\delta f$ is the \Efield-induced shift in the spin transition frequency and $t_E$ is the duration of the electric field pulse.\cite{Mims, EFC}

\medskip

\noindent Representative data are shown in Fig.~\ref{fig: PulseLin}(b), where the integrated echo is plotted as a function of the $E-$field pulse duration. The measurement was conducted at X-band at a temperature of 20 K with an inter-pulse interval $\tau = 2~\mathrm{\mu s}$, under a static magnetic field of $B_0 = 340~\mathrm{mT}$ (corresponding to the ``Int.'' position in Fig.~\ref{fig: Intro}(c)). An electric field of $1.5 \times 10^6~\mathrm{V/m}$ was applied parallel to $\mathbf{B_0}$. The data show a coherent SEC for the photo-excited state of ACRSA, with the in-phase component of the echo signal decreasing as the duration of the $E$-field pulse increases from 0 to $\tau$. The echo signal subsequently recovers as the duration of the $E$-field pulse increases from $\tau$ to $2\tau$, confirming a coherent SEC. On the other hand, the quadrature component of the echo signal remains at zero, independent of the \(E\)-field pulse duration. The lack of $E$-field response in the quadrature channel is due to the combination of a linear SEC and the sample being a randomly oriented spin ensemble \cite{EFC}.

\begin{figure}[H]
    \centering
    \includegraphics[width = \linewidth]{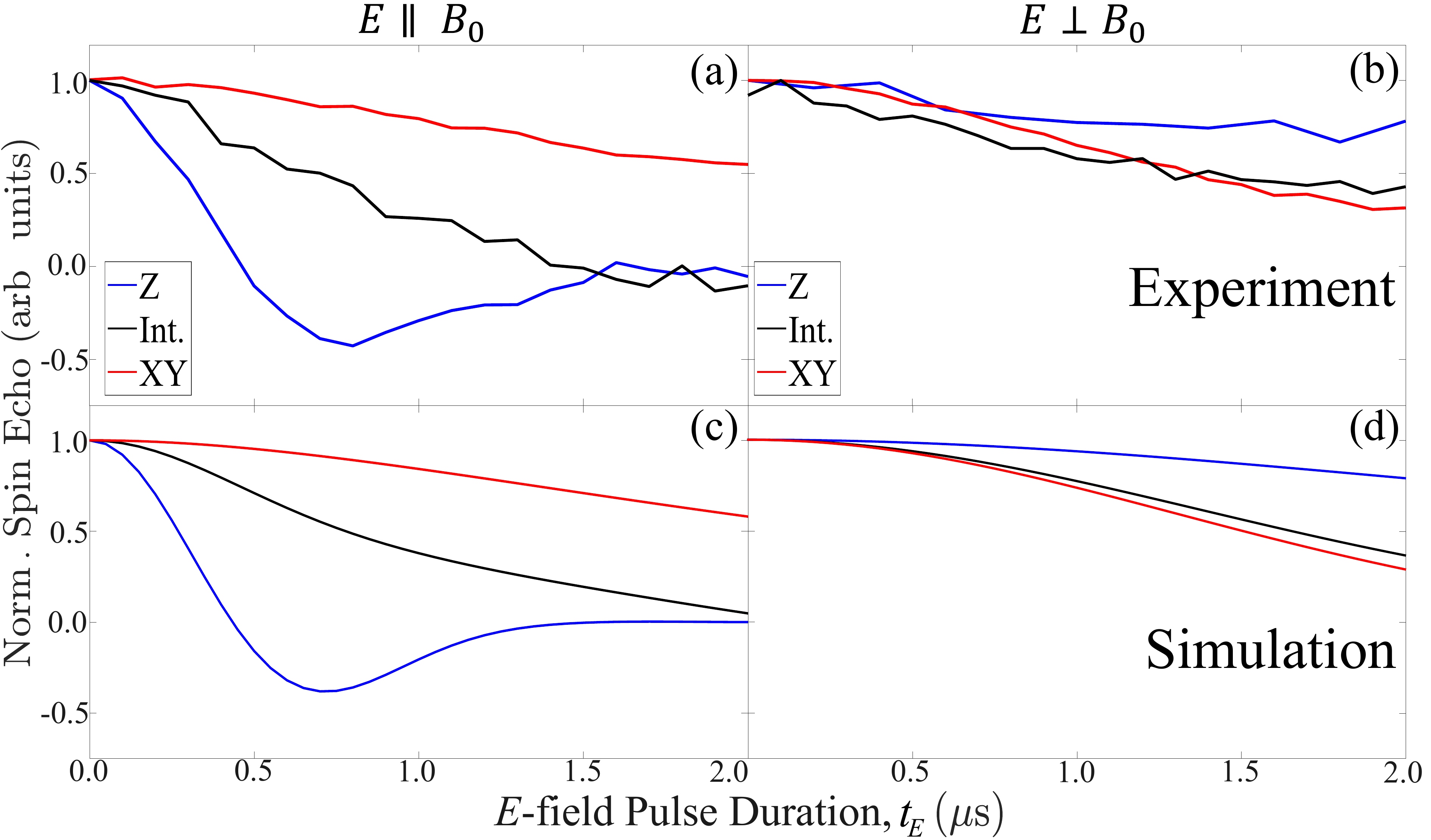}
    \caption{(a-b) Integrated spin-echo intensity as a function of the electric-field pulse duration varied from 0 $\mu$s to $\tau$ (= 2 $\mu$s), for the three field positions considered in the study (see insets). The two configurations correspond to the electric field being parallel (left) and perpendicular (right) to $\mathrm{\mathbf{B_0}}$. In the parallel configuration, the most pronounced electric modulation occurs at Z, while in the perpendicular configuration, it occurs at XY. These positions correspond to the largest alignment between the electric dipole moment/$\mathrm{D}$ and the applied \(E\)-field, supporting the model in which $D$ is the primary contribution to the electric modulation.
    (c-d) Simulations of the electric-field modulation for (a-b), respectively. The model assumes that $\mathrm{D}$ is the only spin Hamiltonian term modulated by the electric field. This modulation is described by $D(E) = D(0) + \kappa E \cdot \cos(\theta)$, where $\mathrm{\theta}$ is the angle between $\mathrm{D}$ and the applied $E$-field. The coupling strength, $\kappa = 0.59 \,\mathrm{Hz/(V/m)}$, quantifies the interaction between the $E$-field and the magnetic anisotropy.
}
    \label{fig: Main Results}

\end{figure}

\medskip

\noindent To verify our hypothesis for the form of the SEC in ACRSA (Eqn.~\ref{eq: dDz}), we measured the $E$-field sensitivity for distinct EPR transitions between well defined quantum states, and we varied the orientation of the \(E\)-field against the molecular orientation \cite{HoW10}. This can be achieved in randomly oriented molecular ensembles by varying the strength of the static magnetic field \( B_0 \), while keeping the EPR frequency fixed. This approach allows for the selection of a subpopulation of molecules with a specific orientation relative to \( \mathbf{B_0} \), due to the molecule’s uniaxial anisotropy \( D \). Fig.~\ref{fig: Intro}(c) shows the three magnetic fields selected for investigation of the SEC, where $B_0$ is (nominally) parallel to (Z), 45$^\circ$ away from (Int.), and perpendicular to (XY) the molecular magnetic anisotropy axis. 

\medskip

\noindent The simulated orientation distribution and the effective electric field for the molecules excited in the EPR experiments at each of these $B_0$ values are shown in Fig. \ref{fig: Distr}(a, b). The latter quantifies how strongly a molecule at a specific orientation \(\theta\) couples to the electric field and it is computed as the projection of the electric field along the \textbf{p}/\(D\) axis ($E \cdot \cos(\theta)$, see inset in panel (b)), weighted by the $\mathbf{B_0}$-dependent molecular population. Such distribution is broadened due to the presence of a sizable $D$ strain ($\sim$150~MHz). Nevertheless, the simulations suggest a good orientation selection in the EPR experiments. The presence of three peaks in the XY and Int. distributions is a consequence of the resonance-field angular dependence in ACRSA, whose details are shown in Fig. S3 in the Supporting Information.

\medskip

\noindent The experimental results are shown in Fig.~\ref{fig: Main Results}(a,b), with simulations based on our simple model shown in Fig.~\ref{fig: Main Results}(c,d). Overall, the SEC effect is strongest when the $E$-field and $\mathrm{\mathbf{B_0}}$ are both parallel to the molecular $D$ axis (Fig.~\ref{fig: Main Results}(a), blue trace); this is the configuration in which the $E$-field couples most strongly to the molecular electric dipole $\mathbf{p}$, and in which the spin transition energies are most sensitive to a change in the magnitude of $D$. The strength of the SEC decreases when the applied \Efield deviates from the molecular dipole moment $\mathbf{p}$, reducing the coupling between the \Efield and $\mathbf{p}$, and when $B_0$ moves away from the anisotropy axis, reducing the sensitivity of the EPR transition to $D$.

\medskip

\noindent The maximum observed $E$-field sensitivity of the EPR transition, $\delta f/E = 0.51 \pm 0.02$ \newline~Hz/(V/m), is comparable to those reported for transition metal-based molecular nanomagnets \cite{EFC, SEC2}. This is despite the spin density residing in the \(\pi\)-system of a molecule with light elements ($p-$orbitals) with negligible atomic SOC, highlighting the importance of the significant molecular electric dipole in facilitating SECs. The results can be explained quantitatively with a SEC coefficient of $\kappa = 0.59\pm 0.03$~Hz/(V/m) (see the Supplementary Information for more fitting details). A strain of this coefficient of $\sigma_{\kappa} = 0.15\pm 0.01$~Hz/(V/m) was also applied in the simulations, which, together with the orientation distribution of the molecules, explain the shapes of the echo intensity vs. $E$-field pulse shown in Fig.~\ref{fig: Main Results}(a) and (c).

\medskip

\noindent We estimate a minimum detectable \Efield of $1.2\times10^5$~V/m (i.e., a minimum $\delta f_\mathrm{min} \simeq 62$~kHz), which would produce a \(\sim\)29\% change in the echo signal with $t_E = \tau = 2$~$\mu$s (as $1 - \cos{(2\pi\delta f_\mathrm{min}\times \tau)} \simeq 0.29$). This detection threshold is limited by the signal-to-noise ratio (SNR), which determines whether a \Efield induced modulation of the spin echo can be resolved. Enhancing the EPR signal, for instance increasing the density of the triplet state, can lead to enhancement in the SNR and better sensitivity. The minimum detectable \Efield can also be improved by increasing $\kappa$ and/or $T_2$. While modifying $\kappa$ requires chemical engineering of the molecule or the use of different SCRPs, the molecular $T_2$ can be enhanced through established techniques such as reducing the environmental nuclear spin bath, for instance by deuterating the host matrix. A longer $T_2$ would lead to the same SNR with a larger $\tau$, hence enabling the same modulation to the spin echo with a smaller \Efield.

\medskip

\noindent The SEC model is further verified by applying the $E$-field perpendicular to the static magnetic field $B_0$. The EPR spectrum is unchanged due to the random orientation nature of the sample. By contrast, the relative alignment between the $E$-field and the molecular electric dipoles is altered in this configuration, leading to different SEC effects. For instance, when measuring the SEC effect with $B_0$ at the Z field position (Fig.~\ref{fig: Main Results}(b)), the $E$-field is mostly perpendicular to the molecular $z$-axis, i.e. to its electric dipole $\mathbf{p}$. Hence, unlike the $E\parallel B_0$ configuration, in which the strongest SEC effect is observed with $B_0$ at the Z field position, the EPR transition is almost insensitive to the $E$-field at the same $B_0$ with $E \perp B_0$. This behaviour is quantitatively reproduced by the simulations shown in Fig.~\ref{fig: Main Results}(d), which employ the same model and parameters as those used for the $E \parallel B_0$ configuration. This provides further support for our SEC model. It is noteworthy that the SEC analysis is not affected by the sign of $D$. Fitting the EPR spectrum and the SEC measurements with $D < 0$ leads to similar conclusions for the SEC (see Fig. S2 in the Supporting Information).

\noindent We note that although the proposed SEC in Eqn.~\ref{eq: dDz} appears to share the similar form as those reported for several single-ion spin systems~\cite{HoW10, SEC2, Vaganov2025, MnZnO}, their underlying mechanisms are substantially different. In ACRSA, $D$ arises from the magnetic dipole interaction between the electron and hole that localise on distinct moieties of the molecule, rather than the atomic SOC associated with the spin-carrying transition-metal or lanthanide atoms. The presence of a substantial $\mathbf{p}$ in ACRSA enables an electrostatic coupling between the electron-hole pair and an external \Efield.\cite{HoW10} The strength of the SEC depends directly on the electric polarisability of the SCRP. However, because the SCRP is delocalised over the molecule, modulation of its electric dipole does not necessarily involve distortions to the geometry of the molecule. This is quite different from the aforementioned systems in which the electric polarisability is strongly associated with the atomic displacements.

\noindent Recent work on Cu(II)-based triangular molecular magnets~\cite{Cini2025} has demonstrated electric-field control of magnetic exchange interactions using EPR spectroscopy. Although it is conceivable that the exchange interaction ($J$) in ACRSA could also respond to an external electric field, the SEC observed in this study is unlikely to originate from such modulation. In ACRSA, $J$ determines the $\sim$30 meV energy gap between the ground singlet and excited triplet states~\cite{Nasu2013}, a transition that lies well beyond the energy scale probed by EPR. Furthermore, unlike the Cu(II)-based triangular molecular magnets, ACRSA does not exhibit a frustrated spin manifold whose degeneracy can be tuned by varying $J$. Thus, the observed SEC is probably dominated by changes in the magnetic anisotropy of the excited triplet state rather than modulation of $J$.

\end{subsection}
\end{section}

\begin{section}{Conclusion and Future Work}

By studying the SEC in a light-induced spin-polarised charge-transfer state in ACRSA, we show that the substantial electric dipole moment associated with the CT state enables coupling between the molecular spin and an external electric field. This finding demonstrates that a sizable SEC can be achieved \textit{via} spin-spin interactions alone, with negligible contribution from atomic SOC. Future work will explore the structure-property relations with the aim of optimizing the SEC response for the design of more sensitive quantum devices.

\medskip

\noindent  The two major constraints to the sensitivity of the current device are the random molecular orientation and the relatively short spin coherence time, which restricts sensing applications to cryogenic temperatures. These limitations could be mitigated by aligning the molecules using magnetic fields \cite{align}, employing magnetic dilution \cite{T2conc}, and deuterating \cite{T2, long_T2} or applying mechanical strains to the host polymer matrix \cite{strainalign}. Together with the capability for optical spin state initialization, such strategies could significantly enhance sensitivity and enable room-temperature operation.

\medskip

\noindent Moreover, a recently proposed class of molecular nanomagnets, termed \textit{molecular colour centres} \cite{MCC1, MCC2} , has attracted significant attention in the molecular quantum information community due to their ability to have their spin states not only initialized but also read out optically—mimicking, for instance, nitrogen-vacancy centres \cite{MCC3} . Some of these compounds have also been theoretically investigated for quantum sensing of \(E\)-fields \cite{E_MCC} . Incorporating our technique and device, which combines microwave, optical, and electric-field excitations, with this class of molecules could further enhance the sensitivity of our sensing scheme, enabling electric-field detection with a small number of molecules.


\end{section}


\begin{subsection}{Experimental Methods.}

\noindent \textit{EPR and Electric-field Equipment.} EPR measurements were carried out using a Bruker Elexsys 580 X-band pulsed spectrometer equipped with a \(^4\)He flow cryostat for temperature control. The sample was housed in a parallel-plate capacitor (see next paragraph) and placed inside a Bruker ER-4118X-MD5-w1 resonator with a 5 mm aperture. Optical excitation of the ACRSA sample was provided by 355 nm laser pulses (10 ns duration, 2 mJ per pulse) from a GWU-Lasertechnik VersaScan Optical Parametric Oscillator, operating at 50 Hz shot repetition time. Electric-field pulses (300 V) were applied using an Avtech AVR-4-B generator, triggered via a Tektronix TDS 210 oscilloscope linked to the spectrometer console and controlled with custom Python scripts.

\medskip

\noindent \textit{Electric-field Device.} The electric field was applied using a parallel-plate capacitor integrated into the EPR resonator, as illustrated in Fig.~S4 of the Supporting Information. Each electrode consisted of two 7.3 mm~\(\times\)~3.5 mm~\(\times\)~0.5 mm quartz chip coated with 100 nm of indium-tin oxide (ITO), which allowed both optical and microwave access to the sample, while minimizing insertion losses of the microwave resonator. T-shaped gold patterns (\(\sim\)150 nm thick) were defined via optical lithography (SUSS MJB4) and deposited by thermal evaporation (Polaron Thermal Evaporator) over a 15 nm copper seed layer. These were wire-bonded (Inseto i-bond 5000) to copper contacts on a 25 mm~\(\times\)~3.5 mm PCB, providing both structural support and electrical connection. The electrodes were terminated above the resonator with a 50~\(\Omega\) load to ensure sharp \(E\)-field pulse edges (15 ns rise/fall time). Plate separation was optimized to balance field strength and sample volume, with 200~\(\mu\)m spacing yielding an effective field of \(1.5 \times 10^6\) V/m at 300 V, across a \(\sim\)3.5~\(\mu\)L frozen sample volume.

\medskip

\noindent \textit{Sample Preparation.} ACRSA was purchased from Ossila (purity 99\%) and dissolved in a solution containing 5\% w/w high-molecular-weight PMMA (Poly(methyl methacrylate), 950k) in anisole (methoxybenzene), resulting in a final ACRSA concentration of 90~$\mu$M. This solution was deposited dropwise onto one of the quartz electrodes of the electric-field device and then spin-coated at 1000~rpm for approximately 60 seconds. Before the film had completely dried, the second quartz electrode was placed on top, allowing the solution to act as an adhesive between the two plates. Upon evaporation of the anisole, a solid ACRSA:PMMA film remained confined between the capacitor plates.

\end{subsection}


\section{Supporting Information}
Outline of the procedure for calculating the SEC constant $\kappa$ and electric-field sensitivity; transient ESR data of ACRSA doped in PMMA at 20~K; simulations of angular distributions and Stark effect coefficients for the case of $D < 0$; angular distributions of the resonances; schematic of the electric-field device; and SEC-modulation simulations for $E \perp B_0$ (PDF).



\section{Supporting Information}

\subsection{\large Estimation of the Spin-Electric Coupling Constant and Sensitivity}\label{sec:sensitivity}

\noindent Here we outline the procedure used to estimate the spin-electric coupling (SEC) constant $\kappa$ (Eq.~2) and the corresponding electric-field sensitivity of ACRSA. As discussed in the main text, the analysis assumes that the zero-field splitting parameter $D$ in Eq.~1 is the only spin Hamiltonian term that is modulated by the external electric field. 

\smallskip

\noindent The estimation of $\kappa$ follows an iterative fitting procedure. Within this framework, $\kappa$ is modelled as a Gaussian-distributed parameter, whose broadening may originate from either the applied $E$-field inhomogeneity or strains to the $D$ parameter. For each trial distribution of $\kappa$, the modulation of the zero-field splitting, $\delta D(E,\theta)$, is computed using Eq.~2 for all values of $\theta$, defined as the angle between the molecular symmetry axis ($z$ axis in the main text) and the static magnetic field $B_0$. The corresponding electric-field-induced shift in resonance frequency, $\delta f(\theta)$, is then calculated as the difference between the eigenvalues of the unperturbed $\hat{\mathcal{H}}(D)$ and the perturbed $\hat{\mathcal{H}}(D + \delta D(E,\theta))$ Hamiltonians.

\smallskip

\noindent The resulting $\theta$-dependent SEC oscillations, $\cos\!\left(2\pi\,\delta f(\theta)\,t_E\right)$, where $t_E$ denotes the duration of the $E$-field pulse, are combined over all orientations to generate the total spin-echo response, with each contribution weighted according to the angular distributions reported in Fig.~3(a) for $E \parallel B_0$ and in Fig.~\ref{fig: perp} for $E \perp B_0$. This weighted summation yields the simulated SEC oscillation patterns, which are then compared with the experimental data. The value of $\kappa$ is iteratively refined until the normalized squared deviation between simulation and experiment falls below a predefined convergence criterion.

\smallskip

\noindent The $\kappa$ value reported in the main text, obtained from this fitting procedure, gives rise to the simulated SEC oscillations shown in Fig.~4(c,d), which closely reproduce the experimental data. This optimized parameter was subsequently used to determine the maximum electric-field sensitivity of the spin sensor, reported as 0.36~Hz/(V/m), defined as the ratio between the mean frequency shift $\delta f$ at the $Z$-field position (Fig.~1(c)) for $E \parallel B_0$ and the applied electric-field amplitude.

\subsection{\large Extended Figures}\label{sec:Sfigures}

\renewcommand{\thefigure}{S\arabic{figure}}  \setcounter{figure}{0}

\begin{figure} \includegraphics[width=\linewidth]{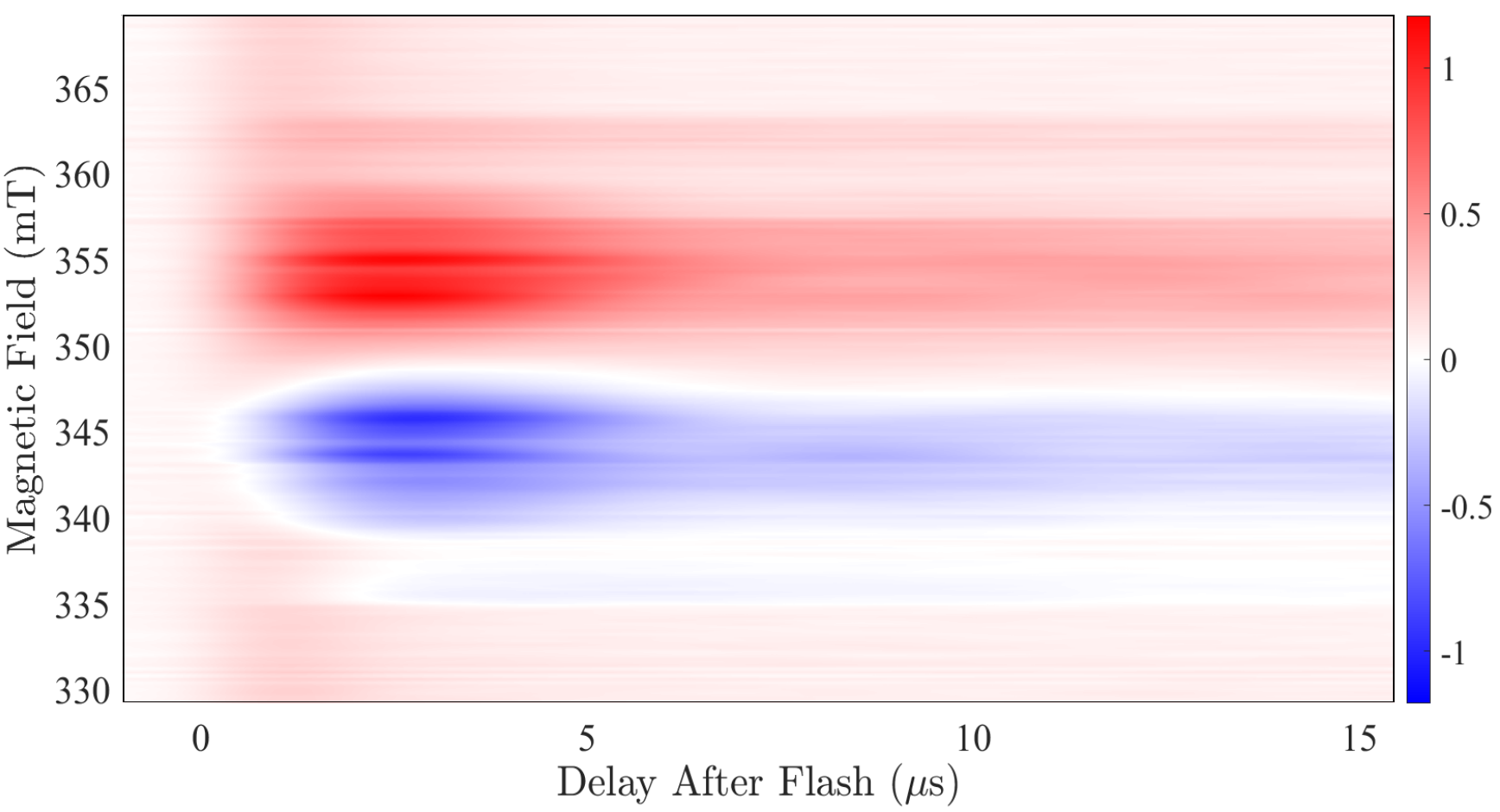} 
 \centering
 \caption{EPR spectrum of a 5\% ACRSA-doped PMMA film measured at 20~K under 355~nm photoexcitation, shown as a function of the time after the laser pulse (known as delay after flash). The data reveal a long-lived triplet state, with the spectral features in Fig. 1(c) persisting for more than 15~\(\mu\)s, i.e., significantly longer than the electron spin-phase memory time at the same temperature (\(\sim 2.5~\mu\)s; see main text). Blue and red areas indicate the emissive and absorptive components of the spectrum, respectively.} 
\label{fig: DAF} 
\end{figure}

\clearpage

\begin{figure}
    \centering
    \includegraphics[width = \linewidth]{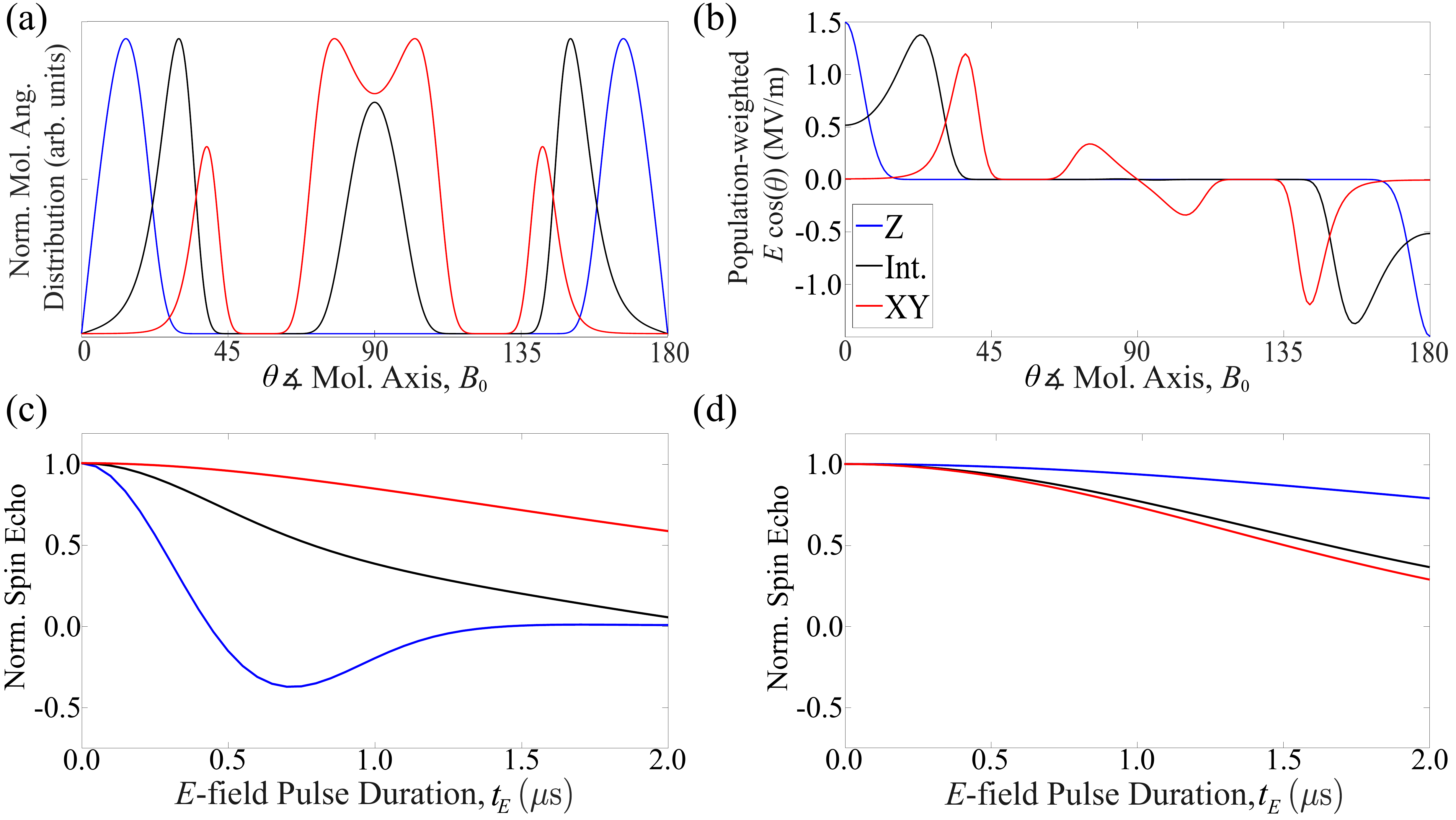}
    \caption{The simulations for (a) the angular distribution, (b) the effective electric coupling, and the electric spin-echo modulations for (c) \( E \parallel B_0 \) and (d) \( E \perp B_0 \) were repeated with a negative magnetic anisotropy \( D \). These simulations yielded results indistinguishable from those presented in the main text, indicating that the spin-electric-coupling model is insensitive to the sign of \( D \).
    }
    \label{fig: dneg}

\end{figure}

\clearpage

\begin{figure}
    \centering
    \includegraphics[width = \linewidth]{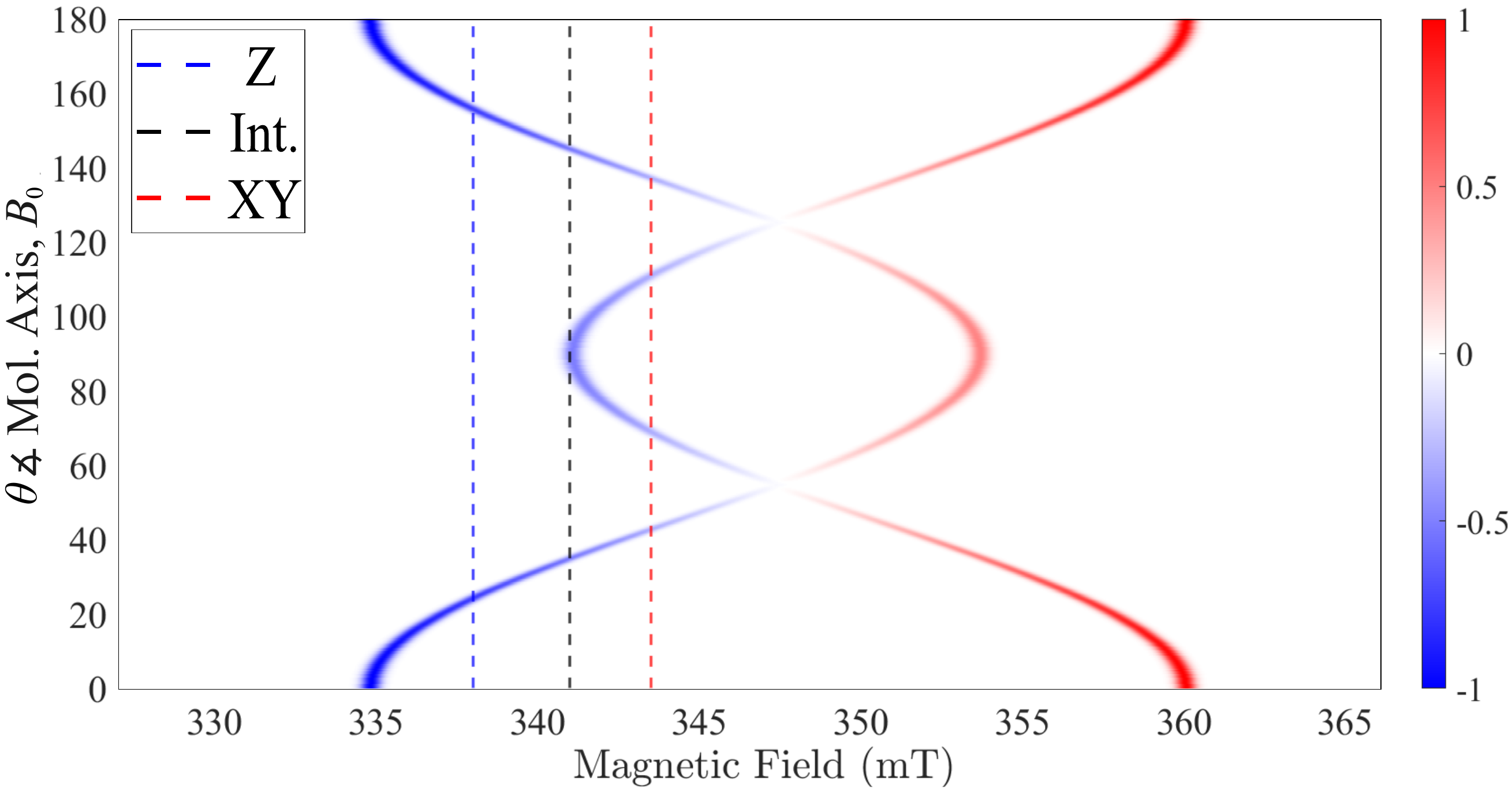}
    \caption{Simulated resonance fields for EPR emission (blue) and absorption (red) transitions as a function of the angle $\theta$ between the molecular axis and the static magnetic field $B_0$. The three vertical dashed lines indicate the specific field positions where we performed spin-electric coupling measurements in ACRSA. By extracting cross-sections at these field values, we obtained the simulated molecular-orientation distributions shown in Fig. 2(a) in the main text, which in turn explain the presence of the three-peak structure observed in the angular distributions at the XY and Int. field positions.
    }
    \label{fig: distr}
\end{figure}

\clearpage

\begin{figure}
    \centering
    \includegraphics[height = 12.0cm, width = 0.7\linewidth]{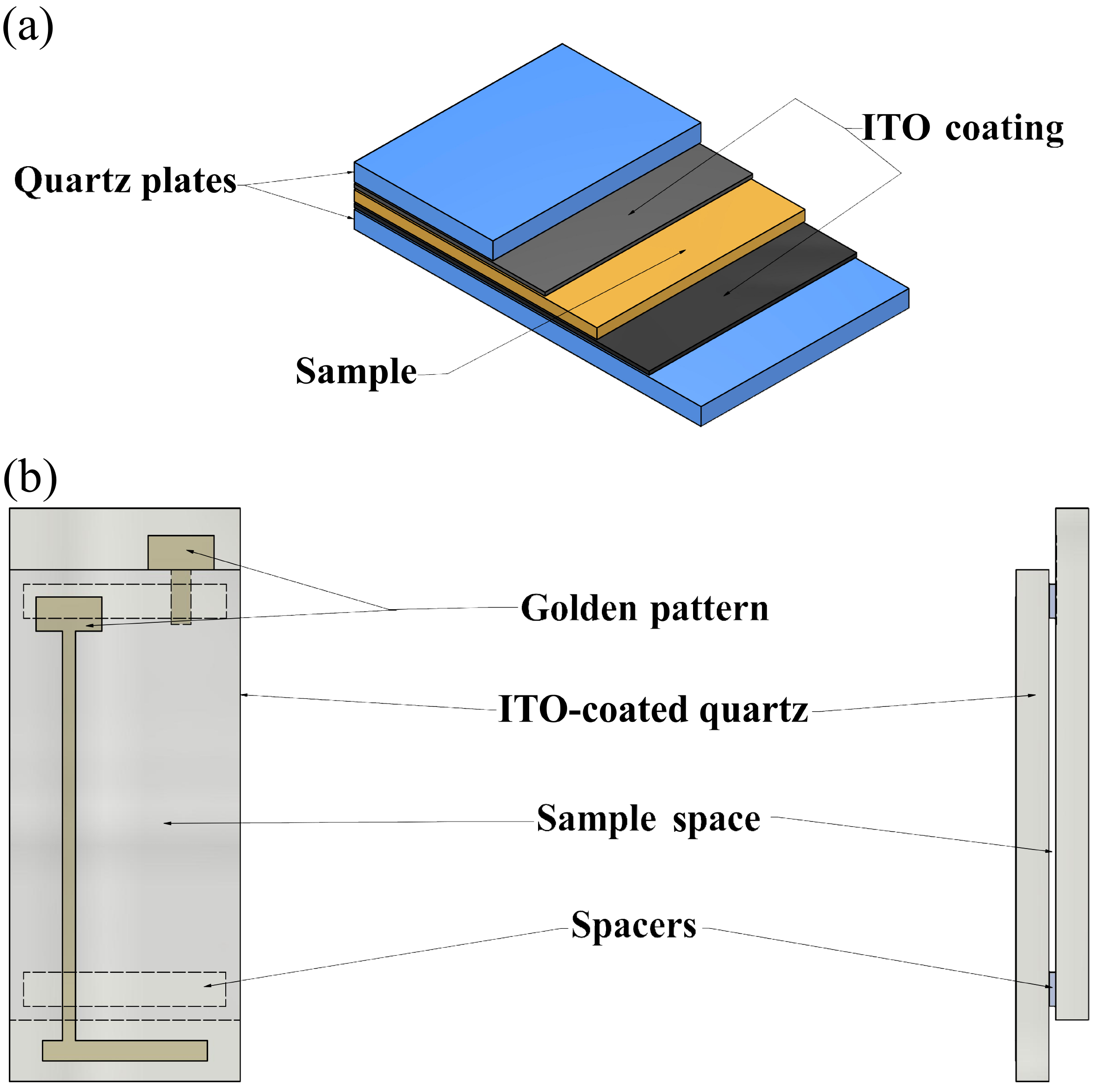}
    \caption{(a) A false-colour scheme of the \(E\)-field device used in the measurements. The DC electric-field pulse is generated by applying a voltage across a parallel-plate capacitor, whose electrodes are made from quartz coated with indium-tin oxide. These electrodes are nearly perfectly transparent to UV and microwave excitations, thus allowing both the initial laser and EPR pulses to interact with the sample. (b) A more detailed schematic of the device, illustrating the T-shaped golden patterns and the offset between the two plates. These features ensure multiple wire bondings between the device and the \(E\)-field generator. With the pulse intensity set to 300 V and the separation between the two capacitor plates roughly equal to 200 \(\mu\)m, an electric field of approximately \(\sim\) 10\(^6\) V/m is generated at the sample site.
    }
    \label{fig: dev}
\end{figure}

\clearpage

\begin{figure}
    \centering
    \includegraphics[ width = \linewidth]{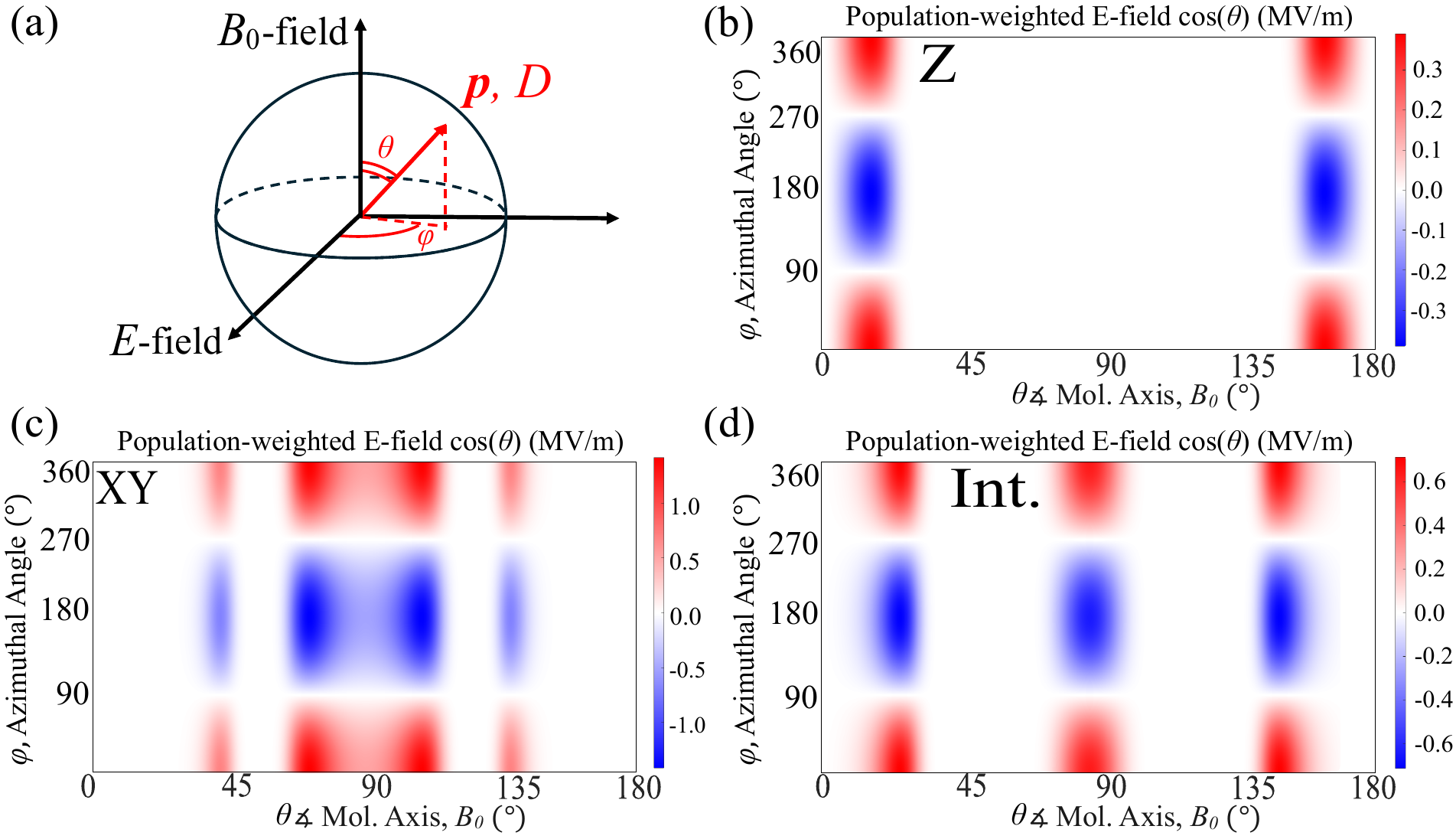}
    \caption{Schematic illustration of the molecular orientation (red arrow), with the molecular axis labelled as ``\(D, \mathbf{p}\)'', in the configuration \(E \perp B_0\). In contrast to the \(E \parallel B_0\) case discussed in the main text, the angle between the \(E\)-field and the molecular axis (determining the strength of the spin-electric coupling) depends on both the polar angle \(\theta\) between the molecular axis and \(B_0\), and the azimuthal angle \(\varphi\) between the plane spanned by \{\(E\), \(B_0\)\} and the molecular axis. (b-d) Angular dependence of the population-weighted effective \(E-\)field (in MV/m) for three field positions: Z (b), XY (c), and Int. (d).
    }
    \label{fig: perp}
\end{figure}

\end{document}